\documentclass[iop]{emulateapj}
\usepackage{color}
\usepackage{graphicx}
\usepackage{amsmath}

\shorttitle{Test of PPN gravity with galaxy-scale strong lensing
systems} \shortauthors{Cao et al.}

\newcommand{\avg}[1]{\langle #1 \rangle}

\begin{document}
\title{ Test of parametrized post-Newtonian gravity with galaxy-scale strong lensing systems }
\author{Shuo Cao\altaffilmark{1}, Xiaolei Li\altaffilmark{1}, Marek Biesiada\altaffilmark{1,2}, Tengpeng Xu\altaffilmark{1}, Yongzhi Cai\altaffilmark{1}, and Zong-Hong Zhu\altaffilmark{1*}
}

\altaffiltext{1}{Department of Astronomy, Beijing Normal University,
100875, Beijing, China; \emph{zhuzh@bnu.edu.cn}}
\altaffiltext{2}{Department of Astrophysics and Cosmology, Institute
of Physics, University of Silesia, Uniwersytecka 4, 40-007 Katowice,
Poland}

\begin{abstract}

Based on a mass-selected sample of galaxy-scale strong gravitational
lenses from the SLACS, BELLS, LSD and SL2S surveys and using a
well-motivated fiducial set of lens-galaxy parameters we tested the
weak-field metric on kiloparsec scales and found a constraint on the
post-Newtonian parameter $\gamma = 0.995^{+0.037}_{-0.047}$ under
the assumption of a flat $\Lambda$CDM universe with parameters taken
from \textit{Planck} observations. General relativity (GR) predicts
exactly $\gamma=1$. Uncertainties concerning the total mass density
profile, anisotropy of the velocity dispersion and the shape of the
light-profile combine to systematic uncertainties of $\sim 25\%$. By
applying a cosmological model independent method to the simulated
future LSST data, we found a significant degeneracy between the PPN
$\gamma$ parameter and spatial curvature of the Universe. Setting a
prior on the cosmic curvature parameter $-0.007< \Omega_k <0.006$,
we obtained the following constraint on the PPN parameter:
$\gamma=1.000^{+0.0023}_{-0.0025}$. We conclude that strong-lensing
systems with measured stellar velocity dispersions may serve as
another important probe to investigate validity of the GR, if the
mass-dynamical structure of the lensing galaxies is accurately
constrained in the future lens surveys.

\end{abstract}


\section{Introduction}\label{sec:introduction}
%%%%%%%%%%%%%%%%%%%%%%%%%%%%%%%%%%%%%%%%%%%%%%%%%%%%%%%%%%%%%%%%%%%%%%%%%%%%%%
%%%%%%%%%%%%%%%%%%%%%%%%%%%%%%%%%%%%%%%%%%%%%%%%%%%%%%%%%%%%%%%%%%%%%%%%%%%%%%

As a successful geometric theory of gravitation,
Einstein's theory of general relativity (GR) has been confirmed in
all observations devoted to its testing to date \citep{Ashby02,Bertotti03}, in particular in famous experiments
\citep{Dyson20,Pound60,Shapiro64,Taylor79}.
However, the pursuit of testing gravity at much higher precision has
ever continued in the past decades, including measurements of the
Earth-Moon separation as a function of time through lunar laser
ranging \citep{Williams04}. On the other hand, formulating and
quantitatively interpreting the test of gravity is another question and an interesting
proposal in this respect has been formulated
in the frameworks of
the parameterized post-Newtonian (PPN) framework \citep{Thorne71}.
Different from the original physical indication \citep{Bertotti03},
scale independent post-Newtonian parameter denoted by $\gamma$, with  $\gamma = 1$ representing GR, may serve as a test
of the theory on large distances.

This paper is focused on the quantitative constraints of the GR as a
theory of gravity, using the recently-released large sample of
galaxy scale strong gravitational lensing systems discovered and
observed in SLACS, BELLS, LSD, and SL2S surveys \citep{Cao15}. Up to
now, most of the progress in strong gravitational lensing has been
made in investigating cosmological parameters
\citep{Zhu00a,Zhu00b,Cha03,Cha04,Mit05,Grillo08,Oguri08,Zhu08a,Zhu08b,Cao12a,Cao12b,Cao12c,Cao12d,Biesiada06,Biesiada10,Collett14,Cardone
et al.2016, Bonvin16}, the distribution of matter in massive
galaxies acting as lenses
\citep{Zhu97,MS98,Jin00,Kee01,Ofek03,Treu06a}, and the photometric
properties of background sources at cosmological distances
\citep{Cao15b}. All the above mentioned results have been obtained
under the assumption that GR is valid. Using strong lensing systems
\citet{Grillo08} reported the value for the present-day matter
density $\Omega_m$ ranging from 0.2 to 0.3 at 99\% confidence level.
This initial result, confirmed in later strong lensing studies (e.g.
\citet{Cao12c, Cao15}) is consistent with most of the current data
including precision measurements of Type Ia supernovae
\citep{Amanullah10} and the anisotropies in the cosmic microwave
background radiation \citep{Planck1}. Currently, the concordance
$\Lambda$CDM model is in agreement with most of the available
cosmological observations, in which the cosmological constant
contributing more than 70\% to the total energy of the universe is
playing the role of an exotic component called dark energy
responsible for accelerated expansion of the Universe. However,
there appeared noticeable tensions between different cosmological
probes. For example, regarding the $H_0$ there is a tension between
the CMB results from Planck \citep{Planck1} and the most recent Type
Ia supernovae data \citep{Riess16}. Similarly, the $\sigma_8$
parameter derived from the CMB results from Planck \citep{Planck1}
turned out to be in tension with the recent tomographic cosmic shear
results both from the Canada France Hawaii Telescope Lensing Survey
(CFHTLenS) \citep{Heymans12,MacCrann15} and the Kilo Degree Survey
(KiDS) \citep{Hildebrandt16}. These tensions partly motivate the
test of GR performed in the present paper.

With reasonable prior assumptions and independent
measurements concerning background cosmology and internal structure of lensing galaxies,
one can use strong lening systems as another tool to constrain the PPN parameters describing the
deviations from the GR. This idea was first adopted on 15 SLACS lenses by
\citet{Bolton06}, who found the post-Newtonian parameter to be
$\gamma=0.98 \pm 0.07$ based on priors on galaxy structure from
local observations. More recently, \citet{Schwab10} re-examined the
expanded SLACS sample \citep{Bolton08a} and obtained a constraint on
the PPN parameter $\gamma = 1.01 \pm 0.05$.

Having available reasonable catalogs of strong lenses: containing
more than 100 lenses, with spectroscopic as well as astrometric data
obtained with well defined selection criteria \citep{Cao15}, the
purpose of this work is to use a mass-selected sample of 80
early-type lenses compiled from SLACS, BELLS, LSD, and SL2S to
provide independent constraints on the post-Newtonian parameter
$\gamma$. Throughout this paper we assume a flat $\Lambda$CDM
cosmology with parameters based on the recent \textit{Planck}
observations \citep{Planck1}.

\section{Method and data}\label{sec:data}

Our goal will be to constrain deviations from General Relativity at the level of $\gamma$ post-Newtonian parameter.
The PPN form of the Schwarzschild metric can be written as
\begin{equation}
d\tau^2 = c^2 dt^2 (1-2GM/c^2r) - dr^2 (1-2\gamma GM/c^2r) - r^2 d\Omega^2
\end{equation}
General Relativity corresponds to $\gamma=1$.

From the theory of gravitational lensing \citep{Schneider92}, for a specific strong
lensing system with the intervening galaxy acting as a lens,
multiple images can form with angular separations close to the so-called Einstein radius
$\theta_E$:
\begin{equation}
\theta_E =  \sqrt{\frac{1+\gamma}{2}} \left(\frac{4G M_E}{c^2}
\frac{D_{ls}}{D_s D_l} \right)^{1/2} ~~~.
\end{equation}
where $M_E$ is the mass enclosed within a
cylinder of radius equal to the Einstein radius, $D_s$ is the distance to the source, $D_l$ is
the distance to the lens, and $D_{ls}$ is the distance between the
lens and the source. All the above mentioned distances are angular-diameter distances.
Rearranging terms with $R_E = D_l \theta_E$
($R$ is the cylindrical radius perpendicular to the
line of sight -- the $\mathcal{Z}$-axis), we obtain a useful formula:
\begin{equation}
\frac{G M_E}{R_E} = \frac{2}{(1+\gamma)} \frac{c^2}{4}
\frac{D_s}{D_{ls}} \theta_E~~~, \label{eq:einrad}
\end{equation}
which indicates that only the matter within the Einstein ring is
important according to the Gauss's law.

On the other hand, spectroscopic measurements  of
central velocity dispersions $\sigma$ in elliptical galaxies, can provide a
dynamical estimate of this mass, based on power-law density profiles for the total mass
density, $\rho$, and luminosity density, $\nu$ \citep{Koopmans06}:
\begin{eqnarray}
\label{eq:rhopl}
\rho(r) &=& \rho_0 \left(\frac{r}{r_0}\right)^{-\alpha} \\
\nu(r) &=& \nu_0 \left(\frac{r}{r_0}\right)^{-\delta}
\label{eq:nupl}
\end{eqnarray}
Here $r$ is the spherical radial coordinate from the lens center: $r^2 = R^2 + \mathcal{Z}^2$. In order to
characterize anisotropic distribution of three-dimensional
velocity dispersion pattern, one introduces \citep{Bolton06,Koopmans06} an anisotropy parameter $\beta$
\begin{equation}
\beta(r) = 1 - {\sigma^2_t} / {\sigma^2_r} \label{eq:beta}
\end{equation}
where $\sigma^2_t$ and $\sigma^2_r$ are, respectively, the tangential
and radial components of the velocity dispersion. In the current analysis we will consider
anisotropic distribution $\beta \neq 0$ and assume, as it almost always is assumed, that $\beta$ is independent of $r$.

Following the well-known spherical Jeans equation \citep{Binney80},
the radial velocity dispersion of the luminous matter $\sigma_r^2(r)$
in the early-type lens galaxies can be expressed as
\begin{equation}
\sigma^2_r(r) =  \frac{G\int_r^\infty dr' \ \nu(r') M(r') (r')^{2
\beta - 2} }{r^{2\beta} \nu(r)}~~~, \label{eq:binney}
\end{equation}
where $\beta$ is a constant velocity anisotropy parameter. Combining
the mass density profiles in Eq.~(\ref{eq:rhopl}), we obtain the
relation between the mass enclosed within a spherical radius $r$ and
$M_E$ as
\begin{equation}
M(r) = \frac{2}{\sqrt{\pi} \lambda(\alpha)}
\left(\frac{r}{R_E}\right)^{3 - \alpha} M_E ~~~,
\end{equation}
where by $\lambda(x) =
\Gamma \left(\tfrac{x-1}{2}\right) / \Gamma
\left(\tfrac{x}{2}\right)$ we denoted the ratio of respective Euler's gamma functions.
Simplifying the formulae with the notation:
$\xi = \delta + \alpha - 2$ taken after
\citep{Koopmans06}, we obtain a convenient form for the radial
velocity dispersion by scaling the dynamical mass to the Einstein
radius:
\begin{equation}
\sigma^2_r(r) = \left[\frac{G M_E}{R_E} \right]
\frac{2}{\sqrt{\pi}\left(\xi- 2 \beta \right) \lambda(\alpha)}
\left(\frac{r}{R_E}\right)^{2 - \alpha}
\end{equation}

In all strong lensing measurements we use, the {\em
observed} velocity dispersion is reported, which is a projected, luminosity
weighted average of the radially-dependent velocity dispersion
profile of the lensing galaxy. Its theoretical value can be calculated
from the Eq.~(\ref{eq:binney}) with the assumption that the relationship between stellar number
density and stellar luminosity density is spatially constant. This
assumption is unlikely to be violated appreciably within the effective
radius of the early-type lens galaxies under consideration.

Moreover, the actual observed velocity dispersion is measured over
the effective spectrometer aperture $\theta_{ap}$ and effectively
averaged by line-of-sight luminosity. Taking into account the
effects of aperture with atmospheric blurring and
luminosity-weighted averaging, the averaged observed velocity
dispersion takes the form
\begin{eqnarray}
\nonumber \bar {\sigma}_*^2 &=& \left[\frac{c^2}{4}
\frac{D_s}{D_{ls}} \theta_E \right] \frac{2}{\sqrt{\pi}}
\frac{(2 \tilde{\sigma}_{\rm atm}^2/\theta_E^2)^{1-\alpha/2}}{ (\xi - 2\beta)} \\
&&\times\left[\frac{\lambda(\xi) - \beta \lambda(\xi+2)}
{\lambda(\alpha)\lambda(\delta)}\right] \frac{
\Gamma(\tfrac{3-\xi}{2}) }{\Gamma(\tfrac{3 - \delta}{2}) } ~~~.
\label{eq:plsig}
\end{eqnarray}
where $\tilde{\sigma}_{\rm atm}\approx\sigma_{\rm atm} \sqrt{1 +
\chi^2 / 4 + \chi^4 / 40}$ and $\chi = \theta_{\rm ap} / \sigma_{\rm
atm}$ \citep{Schwab10}. $\sigma_{\rm atm}$ is the seeing recorded by
the spectroscopic guide cameras during observing sessions
\citep{Cao16}. The above equation tells us that we can constrain the
PPN parameter $\gamma$ on a sample of lenses with known redshifts of
the lens and of the source, with measured velocity dispersion and
the Einstein radius, provided we have reliable knowledge about
cosmological model and about parameters describing the mass
distribution of lensing galaxies ($\alpha$, $\beta$, $\delta$).

For the purpose of our analysis, the angular diameter distances
$D_A(z)$ between reshifts $z_1$ and $z_2$ were calculated using the
best-fit matter density parameter $\Omega_m$ given by
\textit{Planck} Collaboration assuming flat FRW metric
\citep{Planck1}. Moreover, we allow the luminosity density profile
to be different from the total-mass density profile, i.e., $\alpha
\neq \delta$, and the stellar velocity anisotropy exits, i.e.,
$\beta \neq 0$. Based on a well-studied sample of nearby elliptical
galaxies from \citet{Gerhard01}, the anisotropy $\beta$ is
characterized by a Gaussian distribution, $\beta=0.18\pm0.13$, which
is also extensively used in the previous works
\citep{Bolton06,Schwab10}. More recently, \citet{Xu16} measured the
stellar velocity anisotropy parameter $\beta$ and its correlations
with redshifts and stellar velocity dispersion, based on the
Illustris simulated early-type galaxies with spherically symmetric
density distributions. It is worth noting from their results that
$\beta$ markedly depends on stellar velocity dispersion and its mean
value varies from 0.10 to 0.30 for intermediate-mass galaxies (
$200km/s< \sigma_{ap} \leq 300km/s$), which is consistent with the
values used in our analysis.

\begin{figure}
\begin{center}
\includegraphics[width=1.0\hsize]{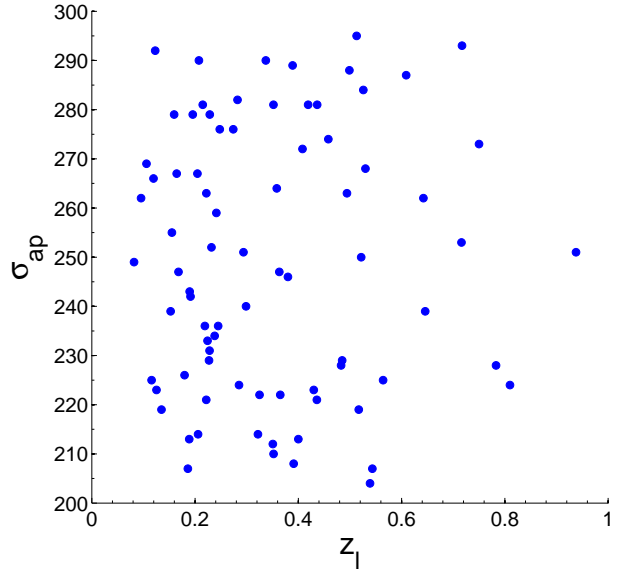}
\end{center}
\caption{ Characteristics of the strong lensing data sample of 80 intermediate mass
early-type galaxies. Observed velocity dispersion inside the aperture is plotted against redshift to the lens.
\label{fig1}}
\end{figure}

Following our previous analysis \citep{Cao16} concerning power-law
mass and luminosity density profiles of elliptical galaxies, we used
a mass-selected sample of strong lensing systems, taken from a
comprehensive compilation of strong lensing systems observed by four
surveys: SLACS, BELLS, LSD and SL2S. The sample has been defined by
restricting the velocity dispersions of lensing galaxies \textbf{to}
the intermediate range: $200km/s< \sigma_{ap} \leq 300km/s$. Lenses
of this sub-sample are located at redshifts ranging from $z_l=0.08$
to $z_l=0.94$. Original data about these strong lenses were derived
by
\citet{Bolton08a,Auger09,Brownstein12,Koopmans02,Treu02,Treu04,Sonnenfeld13a,Sonnenfeld13},
and more comprehensive data concerning these systems can be found in
Table 1 of \citet{Cao15}. Fig.~\ref{fig1} shows the scatter plot for
this sample in the plane spanned by the redshift of the lens and its
velocity dispersion.

\section{Main results}\label{sec:results}

\begin{figure}
\begin{center}
\includegraphics[width=1.0\hsize]{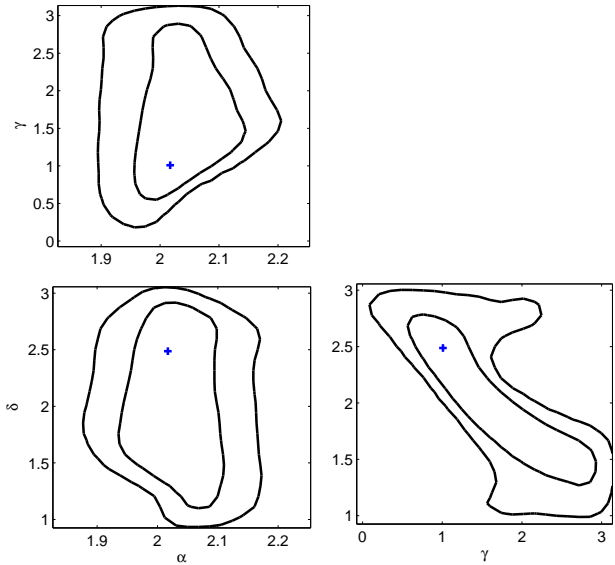}
\end{center}
\caption{ Constraints on the PPN $\gamma$ parameter, the total-mass
and luminosity density parameters obtained from the sample of strong
lensing systems. Blue crosses denote the best-fit values.
\label{fig2}}
\end{figure}

Because $\alpha$, and $\delta$ could not be independently measured
for each lensing galaxy, we firstly treated them as free parameters
and inferred $\alpha$, $\delta$, $\gamma$, simultaneously.
Performing fits on the strong lensing data-set, the 68\% confidence
level uncertainties on the three model parameters are
\begin{eqnarray}
&& \alpha= 2.017^{+0.093}_{-0.082}, \nonumber\\
&& \delta= 2.485^{+0.445}_{-1.393}, \nonumber\\
&& \gamma= 1.010^{+1.925}_{-0.452}. \nonumber
\end{eqnarray}
Fig.~\ref{fig2} shows these constraints in the parameter space of
$\alpha$, $\delta$, and $\gamma$. It is obvious that fits on
$\alpha$ and $\delta$ are well consistent with the analysis results
of \citet{Bolton06,Grillo08,Schwab10}, which are characterized by
Gaussian distributions:
\begin{equation}
\begin{array}{lclclcl}
\avg{\alpha} &=& 2.00 & ; & \sigma_\alpha & = & 0.08 \\
\avg{\delta} & = & 2.40 & ; & \sigma_\delta & = & 0.11~~~.
\end{array}
\label{eq:alphabeta_pars}
\end{equation}
More importantly, the degeneracy between the two parameters,
$\gamma$ and $\delta$, is apparently indicated by the results
presented in Fig.~\ref{fig2}, i.e., a steeper luminosity-density
profile for the lensing early-type galaxies will lead to a larger
value for the parameterized post-Newtonian parameter. This tendency
could also be seen from the sensitivity analysis shown below.

\begin{figure}
\begin{center}
\includegraphics[width=1.0\hsize]{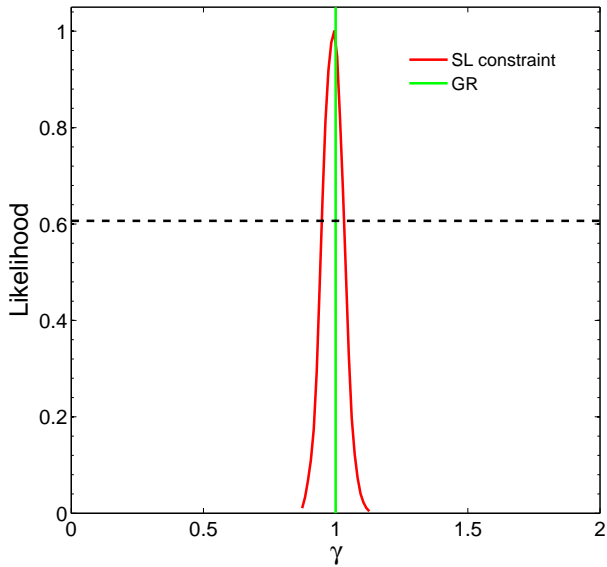}
\end{center}
\caption{ Normalized posterior likelihood of the PPN $\gamma$
parameter obtained with rigid priors on the nuisance parameters
($\alpha$, $\beta$, $\delta$).  \label{fig3}}
\end{figure}

\begin{figure*}
\begin{center}
\includegraphics[width=0.33\hsize]{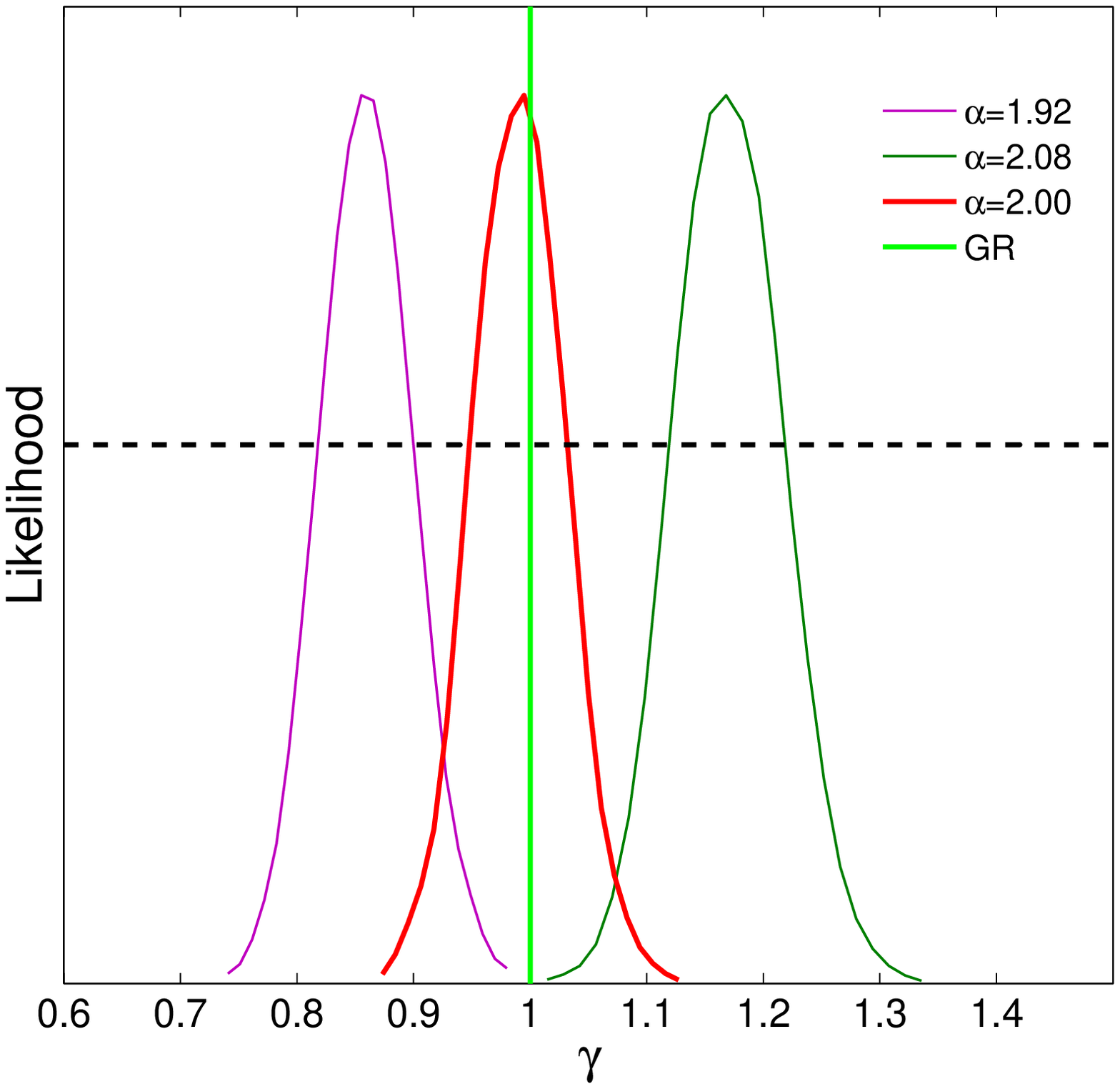} \includegraphics[width=0.33\hsize]{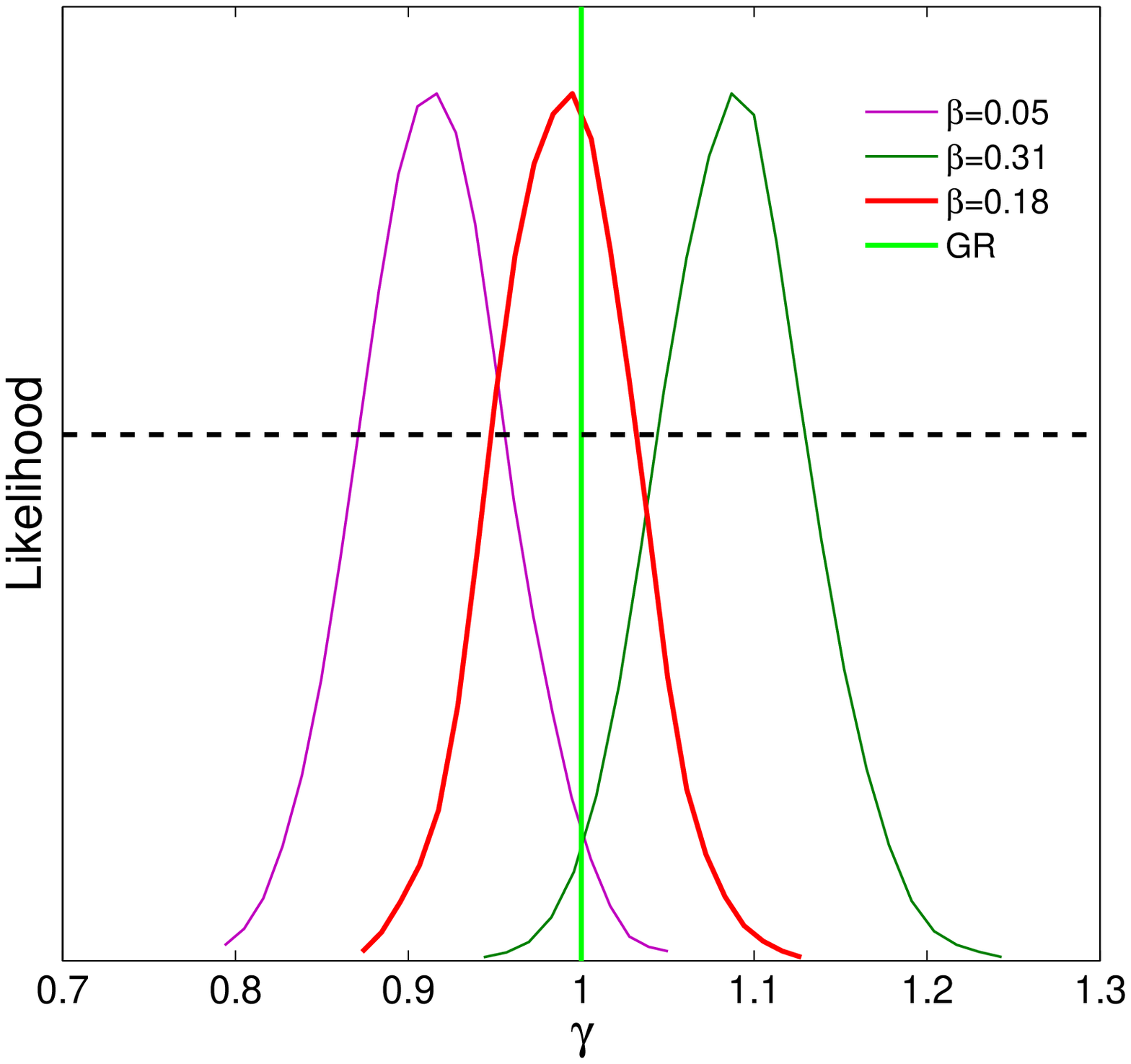}\includegraphics[width=0.33\hsize]{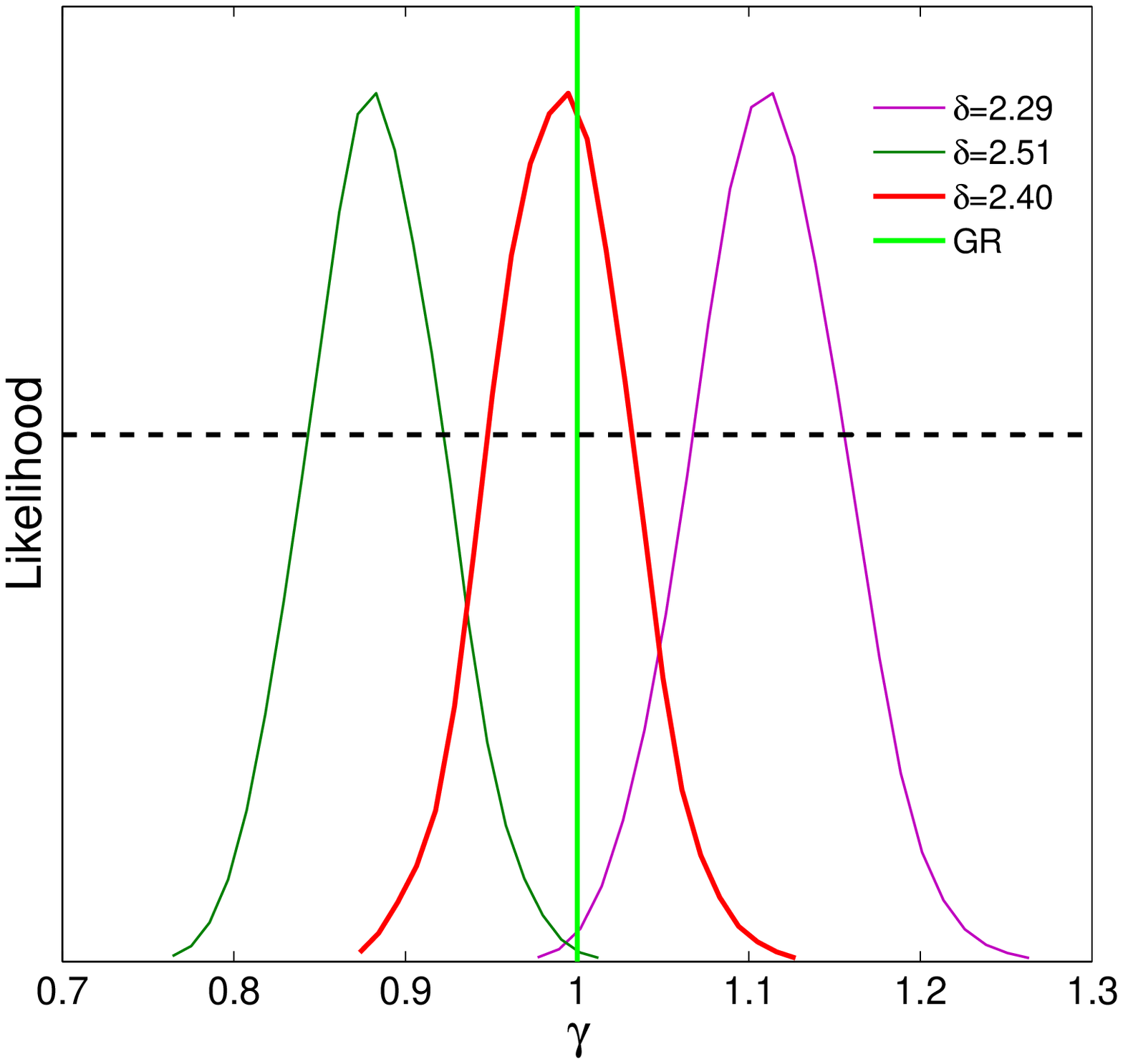}
\end{center}
\caption{Normalized likelihood plot for $\gamma$ by choosing
different galaxy structure parameters. \label{fig7}}
\end{figure*}

Now the parameters characterizing the total mass-profile shape,
velocity anisotropy, and light-profile shape of lenses are set at
their best measured values. Performing fits on $\gamma$, we find the
resulting posterior probability density shown in Fig.~\ref{fig3}.
The result $\gamma = 0.995^{+0.037}_{-0.047}$ (1$\sigma$ confidence)
is consistent with $\gamma = 1$ and also with previous results of
\citep{Bolton06} obtained with strong lensing systems. The scatter
of galaxy structure parameters is an important source of systematic
errors on the final result. Taking the best-fitted values of the
structure parameters as our fiducial model, we investigated how the
PPN constraint is altered by introducing the uncertainties on
$\alpha$, $\beta$, and $\delta$ as listed in
Eq.~(\ref{eq:alphabeta_pars}). Therefore, firstly, we perform a
sensitivity analysis, varying the parameter of interest while fixing
the other parameters at their best-fit values. In general, one can
see from Table~\ref{priorresult} and Fig.~\ref{fig7} that constraint
on $\gamma$ is quite sensitive to small systematic shifts in the
adopted lens-galaxy parameters. By comparing the contribution of
each of these systematic errors to the systematic error on $\gamma$,
we find that the largest sources of systematic error are the mass
density slope $\alpha$, followed by the anisotropy parameter of
velocity dispersion $\beta$ and and the luminosity density slope
$\delta$. Secondly, by considering the intrinsic scatter of
$\alpha$, $\beta$, and $\delta$ into consideration, we found
$\gamma$ varying from 0.845 to 1.240 at 1$\sigma$ confidence level.
It means that systematic errors might exceed $\sim 25\%$ of the
final result. The large covariances of $\gamma$ with $\alpha$ and
$\delta$ seen in Fig.~\ref{fig2} motivate the future use of
auxiliary data to improve constraints on $\alpha$, $\beta$ and
$\delta$. For example, $\alpha$ can be inferred for individual
lenses from high resolution imaging of arcs
\citep{Suyu06,Vegetti10,Collett14,Wong15}, while constraints on
$\beta$ and $\delta$ can be improved with integral field unit (IFU)
data \citep{Barnabe13}, without the assumption of general relativity
(GR).

Another issue which should be discussed is by how much is $\gamma$
affected by the uncertainty of cosmological parameters of the
$\Lambda$CDM model used in our study. For this purpose, we also
considered WMAP9 result of $\Omega_m=0.279$ in order to make
comparison with \textit{Planck} observations. Not surprisingly, the
result was that differences were negligible. It could have been
expected because cosmology intervenes here through the distance
ratio $D_{ls}/D_s$ , which is very weakly dependent on the value of
$\Omega_m$ and in flat cosmology does not depend on $H_0$ at all.

\begin{figure}
\begin{center}
\includegraphics[width=1.0\hsize]{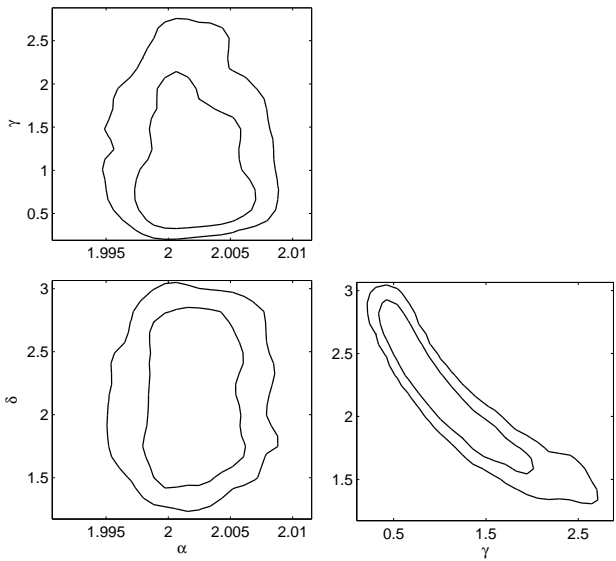}
\end{center}
\caption{ Constraints on the PPN $\gamma$ parameter, the total-mass
and luminosity density parameters obtained from the simulated LSST
strong lensing data. \label{fig8}}
\end{figure}

The next generation wide and deep sky surveys with improved depth,
area and resolution may, in the near future, increase the current
galactic-scale lens sample sizes by orders of magnitude
\citep{Kuhlen04,Marshall05}. Such a significant increase of the
number of strong lensing systems will considerably improve the
constraints on the PPN parameter. Now we will illustrate what kind
of result one could get using the future data from the forthcoming
Large Synoptic Survey Telescope (LSST) survey, which may detect
120000 lenses for the most optimistic scenario \citep{Collett15}. In
order to make a good comparison with the results derived with
current strong lensing systems (Fig.~2), we firstly turn to the
simulated LSST population containing $\sim 40000$ lensing galaxies
with intermediate velocity dispersions ($200km/s< \sigma_{ap} \leq
300km/s$)\footnote{Our simulated LSST sample is obtained with the
simulation programs available on the github.com/tcollett/LensPop.}.
Performing fits on this simulated strong lensing data-set, we obtain
the constraints in the parameter space of $\alpha$, $\delta$, and
$\gamma$ shown in Fig.~\ref{fig8}. It is apparent that from the
simulated LSST strong lensing data, we may expect the total-mass
density parameter $\alpha$ to be estimated with $10^{-3}$ precision.
However, the degeneracy between the PPN $\gamma$ parameter and the
luminosity density parameter $\delta$ still needs to be investigated
with future high-quality integral field unit (IFU) data
\citep{Barnabe13}. In the next section, we will apply a
cosmological-independent method to study the degeneracy
\citep{Rasanen15} between cosmic curvature and parameterized
post-Newtonian parameter $\gamma$.

\begin{table}
\caption{\label{priorresult} Sensitivity of constraints on $\gamma$
with respect to the galaxy structure parameters. }
\begin{center}
\begin{tabular}{c|c}\hline\hline
 Systematics & \hspace{4mm} PPN parameter \hspace{4mm}\\ \hline
$\alpha=2.00;\beta=0.18;\delta=2.40$  & $\gamma = 0.995^{+0.037}_{-0.047}$     \\
$\alpha=1.92;\beta=0.18;\delta=2.40$  & $\gamma= 0.860\pm0.040$     \\
$\alpha=2.08;\beta=0.18;\delta=2.40$  & $\gamma= 1.169\pm0.050$     \\
$\alpha=2.00;\beta=0.05;\delta=2.40$  & $\gamma= 0.914\pm0.043$     \\
$\alpha=2.00;\beta=0.31;\delta=2.40$  & $\gamma= 1.087\pm0.043$     \\
$\alpha=2.00;\beta=0.18;\delta=2.29$  & $\gamma= 1.111\pm0.044$     \\
$\alpha=2.00;\beta=0.18;\delta=2.51$  & $\gamma= 0.883\pm0.039$     \\
 \hline\hline
\end{tabular}
\end{center}

\end{table}

\section{Cosmic curvature and parameterized post-Newtonian formalism}

In a homogeneous and isotropic Universe, the dimensionless distance
$d(z_1;z_s)=(H_0/c)(1+z_s)D_A(z_1;z_s)$ can be written as
\begin{equation}
\label{eq:r} d(z_1;z_s) = \frac{1}{ \sqrt{|\Omega_{k}|}} {\rm
sinn}\left[\sqrt{|\Omega_{k}|}
\int_{z_1}^{z_2}\frac{dz'}{E(z')}\right],
\end{equation}
where $E(z)=H(z)/H_0$ is the expansion rate, and $\Omega_{k}$ is the
spatial curvature density parameter; $sinn(x)= sinh(x)$ for
$\Omega_k>0$, $sinn(x)= x$ for $\Omega_k=0$, and $sinn(x)= sin(x)$
for $\Omega_k<0$, respectively. For a strong lensing system with the
following notation: $d(z)=d(0;z)$, $d_l=d(0; z_l)$, $d_s=d(0; z_s)$,
and $d_{ls}=d(z_l; z_s)$, a simple sum rule could be easily obtained
as
\begin{equation}
\label{dls2}d_{ls}/d_s = \sqrt{1+\Omega_k
d_l^2}-d_l/d_s\sqrt{1+\Omega_k d_s^2}.
\end{equation}
[The case of Eq.~(\ref{dls2}) is given in, e.g., \citet{Peebles03},
p336.] This fundamental formula provides an model-independent probe
to test both the spatial curvature, in combination with weak lensing
and baryon acoustic oscillations (BAO) measurements
\citep{Bernstein06} and the FLRW metric, in combination with strong
lensing systems and SNe Ia observations \citep{Rasanen15}.

For the purpose of our analysis, we determined the dimensionless
distances $d_l$ and $d_s$ of all ``observed'' strong lensing systems
(taken from the LSST simulation by \citet{Collett15}) by fitting a
polynomial to the Union2.1 SN Ia data covering the redshift range
$0<z\leq 1.414$ \citep{Amanullah10}. Therefore we bypassed the need
to assume any specific cosmological model. By using Eq.~(\ref{dls2})
we were able to calculate the distance ratio $d_{ls}/d_s$ depending
only on the curvature density parameter $\Omega_k$. The reported
statistical and systematic uncertainties of the distance modulus for
individual SNe Ia are considered in the fitting procedure. In the
Union2.1 SN Ia compilation, light-curve fitting parameters which are
used for distance estimation are constrained in a global fit.
However, compared to the uncertainties in the modeling of the strong
lensing systems, the model-dependence of the SNe Ia analysis is
likely subdominant \citep{Rasanen15}. Then we assessed the distance
ratios $d_{ls}/d_s$ from the strong lensing data (Einstein radius
and velocity dispersion) according to the Eq.~\ref{eq:plsig}. For
this purpose we used the simulated observations of forthcoming
photometric LSST survey \citep{Collett15}. Using the simulation
programs available on the github.com/tcollett/LensPop, we obtained
53000 strong lensing systems meeting the redshift criterion
$0<z_l<z_s\leq 1.414$ in compliance with SNIa data used in parallel.
The simulated catalog is derived on the base of realistic population
models of elliptical galaxies acting as lenses, with the mass
distribution approximated by the singular isothermal ellipsoids.

\begin{figure}
\begin{center}
\includegraphics[width=1.0\hsize]{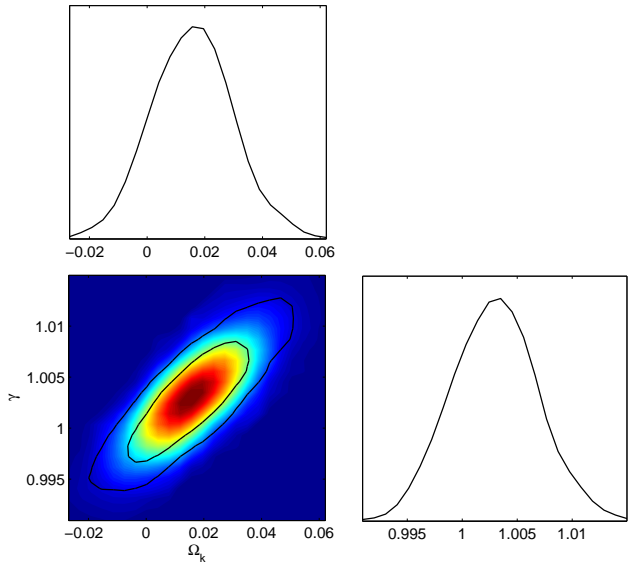}
\end{center}
\caption{ Constraints on the PPN parameter and cosmic curvature from
the simulated LSST strong lensing data. \label{fig4}}
\end{figure}

\begin{figure}
\begin{center}
\includegraphics[width=1.0\hsize]{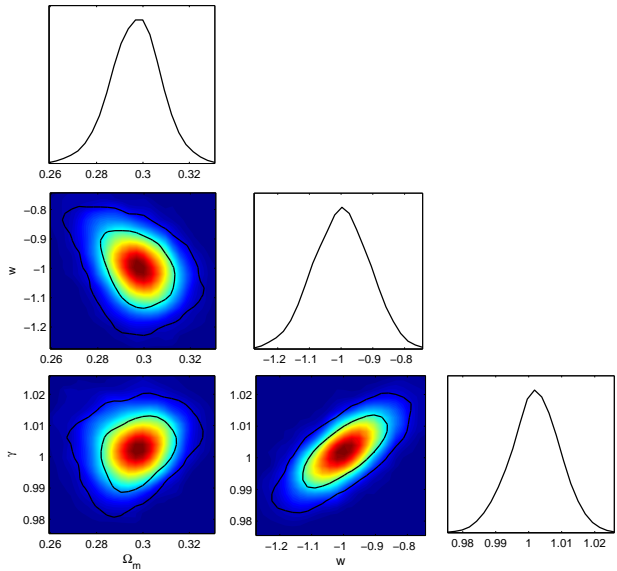}
\end{center}
\caption{ Constraints on the PPN parameter and cosmological
parameters from the simulated LSST strong lensing data.
\label{fig5}}
\end{figure}

Following the assumptions underlying the simulation, we fixed
$\alpha=\delta=2$ and $\beta=0$ in our analysis. We took the
fractional uncertainty of the Einstein radius at the level of 1\%
and the observed velocity dispersion at the level of 10\%. Secondary
lensing contribution from the matter along the line-of-sight was
neglected in our analysis \footnote{ The assumption of 1\% accuracy
on the Einstein radius measurements from future LSST survey is
reasonable, although the line-of-sight effect might introduce $\sim
3\%$ uncertainties in the Einstein radii \citep{Hilbert09}. However,
according to the recent analysis by \citet{Collett16}, the
lines-of-sight for monitorable strong lenses (especially for
quadruply imaged quasars) might be biased at the level of 1\%. Some
attempts to account for the line-of-sight secondary lensing for
quasars can also be found in \citet{Collett13,Greene13,Rusu16}.}.
Fig.~\ref{fig4} displays the fitting results in the
$\Omega_k-\gamma$ plane, thus illustrating the dependence between
the cosmic curvature and the PPN $\gamma$ parameter. It is apparent
that a flat universe together with the validity of GR ($\Omega_k=0$,
$\gamma=1$) is strongly supported. More importantly, it is
interesting to note that there exists a significant degeneracy
between the spatial curvature of the Universe and the PPN parameter,
which captures how much space curvature is provided by unit rest
mass of the objects along or near the path of the particles. Similar
degeneracy between $\gamma$ and the other cosmological parameters
(the matter density fraction, $\Omega_m$ and the equation of state
of dark energy, $w$) can also be seen from Fig.~\ref{fig5}.

\begin{figure}
\begin{center}
\includegraphics[width=1.0\hsize]{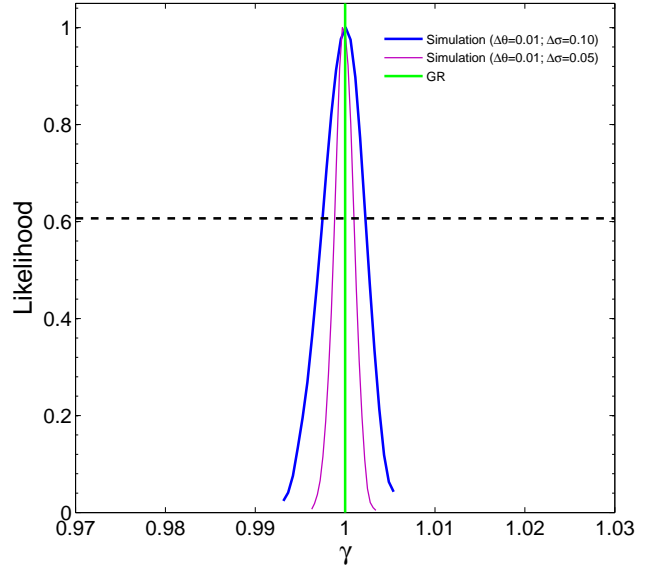}
\end{center}
\caption{ Constraints on the PPN parameter from simulated LSST
strong lensing data, with a prior on the cosmic curvature
$-0.007<\Omega_k<0.006$ from \textit{Planck}. \label{fig6}}
\end{figure}

One can easily check that reduction of the error of $\Omega_k$ would
lead to more stringent fits of $\gamma$, which encourages us to
consider the possibility of testing PPN at much higher accuracy with
future surveys of strong lensing systems. We now set a prior on the
cosmic curvature with $-0.007<\Omega_k<0.006$, according to the
latest CMB data and baryon acoustic oscillation data
\citep{Planck1}, and get a constraint on the PPN parameter:
$\gamma=1.000^{+0.0023}_{-0.0025}$. When we changed the fractional
uncertainty of the Einstein radius to the level of 1\% and the
observed velocity dispersion to the level of 5\%, the resulting
constraint on the PPN parameter became:
$\gamma=1.000^{+0.0009}_{-0.0011}$. The posterior probability
density for $\gamma$ is shown in Fig.~\ref{fig6}. One can see from
this plot that much more severe constraints would be achieved, and
one can expect $\gamma$ to be estimated with $10^{-3}\sim 10^{-4}$
precision.

\section{Conclusions}

Based on a mass-selected galaxy-scale strong gravitational lenes
from the SLACS, BELLS, LSD and SL2S surveys and a well-motivated
fiducial set of lens-galaxy parameters, we tested the weak-field
metric on kiloparsec scales and found a constraint on the
post-Newtonian parameter $\gamma = 0.995^{+0.037}_{-0.047}$ under
the assumption of a flat universe from \textit{Planck} observations.
Therefore it is in agreement with the General Relativity value of
$\gamma=1$ with $4\%$ accuracy. Considering systematic uncertainties
in total mass-profile shape, velocity anisotropy, and light-profile
shape, we estimate systematic errors to be $\sim 25\%$.

Furthermore, we illustrated what kind of result we could get using
the future data from the forthcoming Large Synoptic Survey Telescope
(LSST) survey \citep{Collett15}. We applied a cosmological model
independent method to study the degeneracy \citep{Rasanen15} between
cosmic curvature and the parameterized post-Newtonian parameter
$\gamma$. It is apparent that spatially flat Universe with the
conservation of GR ($\Omega_k=0$, $\gamma=1$) is strongly supported.
Moreover, the reduced uncertainty of $\Omega_k$ leads to more
stringent fits of $\gamma$. This opens up the possibility of testing
PPN with much higher accuracy using strong lensing systems
discovered in the future surveys. By setting a prior on the cosmic
curvature with $-0.007<\Omega_k<0.006$, assumed according to the
latest CMB plus baryon acoustic oscillation data \citep{Planck1},
the accuracy of $\gamma$ determination reached $10^{-3}\sim 10^{-4}$
precision.

Therefore, we conclude that samples of strong lensing systems
with measured stellar velocity dispersions, much larger than currently available, may serve as an
important probe to test the validity of the GR, provided that
mass-dynamical structure of lensing galaxies is better characterized
and constrained in the future surveys.

\vspace{0.5cm}

This work was supported by the Ministry of Science and Technology
National Basic Science Program (Project 973) under Grants Nos.
2012CB821804 and 2014CB845806, the Strategic Priority Research
Program ``The Emergence of Cosmological Structure" of the Chinese
Academy of Sciences (No. XDB09000000), the National Natural Science
Foundation of China under Grants Nos. 11503001, 11373014 and
11073005, the Fundamental Research Funds for the Central
Universities and Scientific Research Foundation of Beijing Normal
University, China Postdoctoral Science Foundation under grant No.
2015T80052, and the Opening Project of Key Laboratory of
Computational Astrophysics, National Astronomical Observatories,
Chinese Academy of Sciences. Part of the research was conducted
within the scope of the HECOLS International Associated Laboratory,
supported in part by the Polish NCN grant DEC-2013/08/M/ST9/00664 -
M.B. gratefully acknowledges this support. M.B. obtained approval of
foreign talent introducing project in China and gained special fund
support of foreign knowledge introducing project.


\begin{thebibliography}{}


\bibitem[Ade et al.(2014)]{Planck1} Ade, P.A.  R., et al. 2014, A\&A, 571, A16
\bibitem[Amanullah et al.(2010)]{Amanullah10} Amanullah, R., et al. 2010, ApJ, 716, 712
\bibitem[Ashby(2002)]{Ashby02} Ashby, N. 2002, Phys. Today, 55, 41
\bibitem[Auger et~al.(2009)]{Auger09} Auger, M. W., et al. 2009, ApJ, 105, 1099
\bibitem[Barnab\`{e} et al.(2013)]{Barnabe13} Barnab\`{e}, M., et al. 2013, MNRAS, 436, 253
\bibitem[Bernstein(2006)]{Bernstein06} Bernstein, G. 2006, ApJ, 637, 598
\bibitem[Bertotti et al.(2003)]{Bertotti03} Bertotti, B., Iess, L., \& Tortora, P. 2003, Nature, 425, 374
\bibitem[Biesiada(2006)]{Biesiada06} Biesiada, M. 2006, PRD, 73, 023006
\bibitem[Biesiada, Pi\'{o}rkowska, \& Malec(2010)]{Biesiada10} Biesiada, M., Pi\'{o}rkowska, A., \& Malec, B. 2010, MNRAS, 406, 1055
\bibitem[Binney(1980)]{Binney80} Binney, J. 1980, MNRAS, 190, 873
\bibitem[Bolton et al.(2006)]{Bolton06} Bolton, A. S., Rappaport, S., \& Burles, S. 2006, PRD, 74, 061501
\bibitem[Bolton et~al.(2008a)]{Bolton08a} Bolton, A. S., et al. 2008, ApJ, 682, 964
\bibitem[Bonvin et~al.(2016)]{Bonvin16} Bonvin, V., et al. 2016, arXiv:1607.01790
\bibitem[Brownstein et~al.(2012)]{Brownstein12} Brownstein, et al. 2012, ApJ, 744, 41
\bibitem[Cardone et al.(2016)]{Cardone et al.2016} Cardone, V. F., Piedipalumbo, E., Scudellaro, P. 2016 MNRAS, 455, 831-837.
\bibitem[Cao \& Zhu(2012)]{Cao12a} Cao, S., \& Zhu, Z.-H. 2012, A\&A, 538, A43
\bibitem[Cao, Covone \& Zhu(2012)]{Cao12b} Cao, S., Covone, G., \& Zhu, Z.-H. 2012, ApJ, 755, 31
\bibitem[Cao et al.(2012)]{Cao12c} Cao, S., Pan, Y., Biesiada, M., Godlowski, W., \& Zhu, Z.-H. 2012, JCAP, 03, 016
\bibitem[Cao, Zhu \& Zhao(2012)]{Cao12d}  Cao, S., Zhu, Z.-H., \& Zhao, R. 2012, PRD, 84, 023005
\bibitem[Cao \& Zhu(2014)]{Cao14}  Cao, S., \& Zhu, Z.-H. 2014, PRD, 90, 083006
\bibitem[Cao et al.(2015a)]{Cao15} Cao, S., et al., 2015a, ApJ, 806, 185
\bibitem[Cao et al.(2015b)]{Cao15b} Cao, S., et al., 2015b, AJ, 149, 3
\bibitem[Cao et al.(2016)]{Cao16} Cao, S., et al., 2016, MNRAS, 461, 2192
\bibitem[Chae(2003)]{Cha03} Chae, K.-H. 2003, MNRAS, 346, 746
\bibitem[Chae et al.(2004)]{Cha04} Chae, K.-H., Chen, G., Ratra, B., \& Lee, D.-W. 2004, ApJ, 607, L71
\bibitem[Collett et al.(2013)]{Collett13} Collett, T. E., et al. 2013, MNRAS, 432, 679
\bibitem[Collett \& Auger(2014)]{Collett14} Collett, T. E. \& Auger, M. W. 2014, MNRAS, 443, 969
\bibitem[Collett(2015)]{Collett15} Collett, T. E. 2015, arXiv:1507.02657
\bibitem[Collett \& Cunnington(2016)]{Collett16} Collett, T. E. \& Cunnington, S. D. 2016, MNRAS, 462, 3255
\bibitem[Dyson et al.(1920)]{Dyson20} Dyson, F. W., Eddington, A. S., \& Davidson, C. 1920, Phil. Trans. R. Soc., 220, 291
\bibitem[Gerhard et al.(2001)]{Gerhard01} Gerhard, O., Kronawitter, A., Saglia, R. P., \& Bender, R. 2001, AJ, 121, 1936
\bibitem[Golse et al.(2002)]{Golse02} Golse, G., Kneib, J.-P., \& Soucail, G. 2002, A\&A, 387, 788
\bibitem[Greene et al.(2013)]{Greene13} Greene, Z. S., et al. 2013, ApJ, 768, 39
\bibitem[Grillo et~al.(2008)]{Grillo08} Grillo, C., Lombardi, M., \& Bertin, G. 2008, A\&A, 477, 397
\bibitem[Heymans et al.(2012)]{Heymans12} Heymans, C., et al. 2012, MNRAS, 427, 146
\bibitem[Hilbert et al.(2009)]{Hilbert09} Hilbert, S., et al. 2009, A\&A, 499, 31
\bibitem[Hildebrandt et al.(2016)]{Hildebrandt16} Hildebrandt, H., et al. 2016, preprint [arXiv:1603.07722]
\bibitem[Jin et al.(2000)]{Jin00} Jin, K.-J., Zhang, Y.-Z., \& Zhu, Z.-H. 2000, PLA, 264,335
\bibitem[Keeton(2001)]{Kee01} Keeton, C.~R. 2001, ApJ, 561, 46
\bibitem[Koopmans \& Treu(2002)]{Koopmans02} Koopmans, L.V.E, \& Treu, T. 2002, ApJ, 583, 606
\bibitem[Koopmans(2006)]{Koopmans06} Koopmans, L.V.E. 2006, in EAS Publications Series, ed. G. A. Mamon, F. Combes, C. Deffayet, \& B. Fort, Vol. 20, 161
\bibitem[Kuhlen et al.(2004)]{Kuhlen04} Kuhlen, M., Keeton, C. R., \& Madau, P. 2004, ApJ, 601, 104
\bibitem[MacCrann et al.(2015)]{MacCrann15} MacCrann, N., Zuntz, J., Bridle, S., Jain, B., Becker, M. R. 2015, MNRAS, 451, 2877
\bibitem[Mao \& Schneider(1998)]{MS98} Mao, S. D., \& Schneider, P. 1998, MNRAS, 295, 587
\bibitem[Marshall et al.(2005)]{Marshall05} Marshall, P., Blandford, R., \& Sako, M. 2005, NAR, 49, 387
\bibitem[Mitchell et al.(2005)]{Mit05} Mitchell, J.~L., Keeton, C.~R., Frieman, J.~A., \& Sheth, R.~K. 2005, ApJ, 622, 81
\bibitem[Ofek et al.(2003)]{Ofek03} Ofek, E. O., Rix, H.-W., \& Maoz, D. 2003, MNRAS, 343, 639
\bibitem[Oguri et~al.(2008)]{Oguri08} Oguri, M., et al. 2008, AJ, 135, 512
\bibitem[Peebles(1993)]{Peebles03} Peebles, P. J. E. 1993, Principles of Physical Cosmology (Princeton University Press, Princeton, NJ, 1993
\bibitem[Pound \& Rebka(1960)]{Pound60} Pound, R. V., \& Rebka, G. A. 1960, PRL, 4, 337
\bibitem[R\"{a}s\"{a}nen et~al.(2015)]{Rasanen15} R\"{a}s\"{a}nen, S., Bolejko, K., \& Finoguenov, A. 2015, PRL, 115, 101301
\bibitem[Riess et al.(2016)]{Riess16} Riess, A. G., et al. 2016, ApJ, 826, 56
\bibitem[Rusu et al.(2016)]{Rusu16} Rusu, C. E., et al. 2016, arXiv:1607.01047
\bibitem[Schneider et al.(1992)]{Schneider92} Schneider, P., Ehlers, J., \& Falco, E. E. 1992, Gravitational Lenses (Springer-Verlag, New York)
\bibitem[Schwab et al.(2010)]{Schwab10} Schwab, J., Bolton, A. S., \& Rappaport, S. A. 2010, ApJ, 708, 750
\bibitem[Shapiro(1964)]{Shapiro64} Shapiro, I. I. 1964, Phys. Rev. Lett., 13, 789
\bibitem[Sonnenfeld et~al.(2013a)]{Sonnenfeld13a} Sonnenfeld, A., Gavazzi, R., Suyu, S.H., Treu, T., Marshall, P.J. 2013a, ApJ, 777, 97 [arXiV:1307.4764]
\bibitem[Sonnenfeld et~al.(2013b)]{Sonnenfeld13} Sonnenfeld, A., Treu, T., Gavazzi, R., Suyu, S.H., Marshall, P.J., Auger, M.W., Nipoti, C., 2013b, ApJ, 777, 98
\bibitem[Suyu et al.(2006)]{Suyu06} Suyu, S. H., et al. 2007 AAS/AAPT Joint Meeting, American Astronomical Society Meeting 209, id.21.02; Bulletin of the American Astronomical Society, Vol. 38, p.927
\bibitem[Taylor et al.(1979)]{Taylor79} Taylor, J. H., Fowler, L. A., \& McCulloch, P. M. 1979, Nature, 277, 437
\bibitem[Thorne \& Will(1971)]{Thorne71} Thorne, K. S., \& Will, C. M. 1971, ApJ, 163, 595
\bibitem[Treu \& Koopmans(2002)]{Treu02} Treu, T., \& Koopmans, L.V.E. 2002, ApJ, 575, 87
\bibitem[Treu \& Koopmans(2004)]{Treu04} Treu, T.,\& Koopmans, L.V.E. 2004, ApJ, 611, 739
\bibitem[Treu et al.(2006a)]{Treu06a} Treu, T., et al. 2006a, ApJ, 640, 662
\bibitem[Vegetti et al.(2010)]{Vegetti10} Vegetti, S., Koopmans, L. V. E., Bolton, A., Treu, T., \& Gavazzi, R. 2010, MNRAS, 408, 1969
\bibitem[Williams et al.(2004)]{Williams04} Williams, J. G., Turyshev, S. G., \& Boggs, D. H. 2004, PRL, 93, 261101
\bibitem[Wong et al.(2015)]{Wong15} Wong, K. C., Suyu, S. H., \& Matsushita, S. 2015, ApJ, 811, 115
\bibitem[Xu et al.(2016)]{Xu16}  Xu, D. D., et al. 2016, arXiv:1610.07605v1
\bibitem[Zhu \& Wu(1997)]{Zhu97} Zhu, Z.-H., \& Wu, X.-P. 1997, A\&A, 324, 483
\bibitem[Zhu(2000a)]{Zhu00a} Zhu, Z.-H. 2000, MPLA, 15, 1023
\bibitem[Zhu(2000b)]{Zhu00b} Zhu, Z.-H. 2000, IJMPD, 9, 591
\bibitem[Zhu \& Sereno(2008a)]{Zhu08a} Zhu, Z.-H., \& Sereno, M. 2008, A\&A, 487, 831
\bibitem[Zhu et al.(2008b)]{Zhu08b} Zhu, Z.-H., et al. 2008, A\&A, 483, 15



\end{thebibliography}
\end{document}